# The hybrid dilemma —do hybrid technologies play a transitionary or stationary role in transitions processes?

Amir Mirzadeh Phirouzabadi

a.m.phirouzabadi@tik.uio.no, amir.mirzadeh@uon.edu.au

Centre for Technology, Innovation & Culture (TIK)

Faculty of Social Sciences, University of Oslo, Norway

## ABSTRACT

This research unravels the stationary or transitionary dilemma of hybrid technologies in transitions processes. A system dynamics technology interaction framework is built and simulated based on Technological Innovation System and Lotka-Volterra models to investigate the inter-technology relationship impacts and modes that hybrid technologies establish with incumbent and emerging technologies. Data collection and simulation are conducted for the case of conventional, hybrid and battery electric vehicles under seven scenarios including landscape pressure, predator-prey, niche-incumbent, hybrid-incumbent, and sociotechnical transition over 1985-2070 in the US market. Results reveal that, by acting as an exploration-hybrid solution, hybrid technologies maintain a transitionary role by supporting mainly the technological development side of emerging technology. On the contrary, by acting as an exploitation-hybrid solution, they hardly (or never) sustain an inhibitive role against both the technological and market development sides of incumbent technology. While hybrid technologies may play a stationary role on the market development side in transitions processes, simulation results show that maintaining all inter-technology relationship modes as business-as-usual (i.e., baseline scenario) but instead simultaneously strengthening the various socio-technical dimensions of emerging technology and destabilising the various socio-technical dimensions of incumbent technology (i.e., sociotechnical scenario) is a more promising pathway in both short term (e.g., an accelerated uptake of emerging technology and decline of incumbent technology) and long term (e.g., highest reduction of accumulative GHG emissions). Findings, additionally, reinforce the existence of both spillover and try-harder versions of 'sailing-ship effect', which are either seriously doubted in the literature or partially validated using raw bibliometric and patents data.

## 1. Introduction

Sectors such as energy, water, transportation, and agriculture demand deep, sustainable transitions from dysfunctional locked-in incumbent technologies, such as internal combustion engine vehicles (ICEVs), to new environmentally friendly technologies, such as battery electric vehicles (BEVs), through transformation processes that fundamentally change the way the stable regime has fulfilled certain societal functions (Köhler et al., 2019; Raven, 2007; Raven & Walrave, 2020). One solution to unlock incumbent technology and accelerate transitions processes towards an emerging technology is developing and deploying a hybridised innovation of both incumbent and emerging technologies (Andersson & Jacobsson, 2000; Bergek et al., 2013; Furr & Snow, 2014; Geels, 2002; Pohl, 2011; Sandén & Hillman, 2011). In general, a hybrid technology is defined as a technology that combines two or more technologies with the aim to achieve a more efficient system, such as hybrid electric vehicles (HEVs), wind-diesel systems, and fuel cell-gas turbine systems.

Technologies may establish various relationships with each other depending on the point of interaction. Pistorius and Utterback (1997) conceptualised interactions between an emerging technology and an incumbent technology for market share using the three biological relationship modes of competition, symbiosis, and parasitism. Competition occurs when two technologies have a negative effect on the market growth of each other; symbiosis occurs when they have a positive effect on the market growth of each other; and parasitism occurs when one technology has a positive effect on the market growth of the other while being negatively affected by it. Some transition studies observed that emerging and incumbent technologies may establish a symbiotic relationship with each other when they are temporarily combined as a hybrid technology, but similar to Pistorius, they did not specify or observe the modes of relationships that hybrid technologies establish with both emerging and incumbent technologies. (Geels, 2002; Geels, 2005 ; Geels et al., 2016). Sandén and Hillman (2011) extended the three relationship modes proposed for inter-technology relationships by Pistorius and Utterback (1997) by including the other biological relationship modes, i.e., commensalism, amensalism and neutralism. Commensalism occurs when there is a positive effect for one technology, yet the other technology is not affected; amensalism occurs when there is a negative effect for one technology, yet the other is not affected; and neutralism occurs when there are no effects on either technology. Accordingly, they defined a hybrid technology as “a bridging technology [that] parasitizes on an established technology, while a third technology parasitizes on the bridging technology.” (Sandén & Hillman, 2011, pp. 407). Similar to Sandén and Hillman (2011), Mirzadeh Phirouzabadi et al. (2020g) discussed that such bridging technologies establish parasitic relationships with both incumbent and emerging technologies not only on market share but also on other dimensions including knowledge development and diffusion, entrepreneurial activities, and resource mobilisation. Pretorius et al. (2015) simulated a bridging technology situation and identified that the full transition can be accelerated and completed if and only if the bridging technology establishes parasitic relationships with both emerging and incumbent technologies on market share. They found out that the transition would oscillate forth and back between emerging and incumbent technologies if the bridging technology establishes a commensal relationship with the incumbent technology and a parasitic relationship with the emerging technology.

There are several empirical studies that observed the relationships that a hybrid technology establishes with both emerging and incumbent technologies. For instance, the following relationship modes were observed between HEVs and BEVs in the automotive industry: a commensal relationship on expectations and social image and technological capability and knowledge, a competitive relationship between them on market share (Dijk, 2014), a mix of commensalism, symbiosis, and parasitism on technological knowledge development (Mirzadeh Phirouzabadi et al., 2020c; Mirzadeh Phirouzabadi et al., 2020e), and a mix of amensalism, neutralism, symbiosis, and parasitism on knowledge diffusion, collaborations, and entrepreneurial activities (Mirzadeh Phirouzabadi et al., 2020a; Mirzadeh Phirouzabadi et al., 2020b; Mirzadeh Phirouzabadi et al., Unpublished manuscript).

The literature review reveals that there are inconsistent observations on the modes of relationships as well as the points of interaction that hybrid technologies establish with both emerging and incumbent technologies. Additionally, the studies did not specify if and under what conditions hybrid technologies may play a transitionary or stationary role in transitions processes. There are two contrasting views on the role of hybrid technologies in transitions processes. The first view, which is an optimistic view, considers hybrid technologies as a temporary and transitionary solution that can provide a learning, forward-looking mechanism that cultivates a managing environment for firms to deal with high uncertainty about the future of emerging technologies and the timing of the transition (Cohen & Tripsas, 2018; Furr & Snow, 2014; Fuzes, 2020). In the sustainability transitions research, for instance, hybrid technologies are defined as bridging technologies that act as a catalyst in the transition by parasitising on incumbent technology while being parasitised by emerging technology, i.e., a predator-prey relationship (Bergek et al., 2013; Geels, 2002; Geels, 2005; Geels et al., 2016; Pretorius et al., 2015; Sandén & Hillman, 2011). In contrast, the conservative view is that hybrid technologies are nothing but a trap to transitions as they are merely a physical manifestation of organisational inertia and firms approach hybrid technologies rather from the perspective of incumbent technology; hence, they are inclined to fall back on their learned knowledge and patterns (Avadikyan & Llerena, 2010; Furr & Snow, 2014; Henderson & Clark, 1990; Rosenbloom, 2000; Suarez et al., 2018; Sull et al., 1997).

These contrasting perspectives create a great dilemma on the agenda of managers and policy makers regarding whether hybrid technologies are really what they are expected to be and whether they can fulfil what they are expected to fulfil in transitions processes. This paper aims to unravel the hybridisation dilemma by investigating the stationary and transitionary roles of hybrid technologies in transitions processes. A system dynamics technology interaction framework is built and simulated based on Technological Innovation System and Lotka-Volterra models to investigate the inter-technology relationship impacts and modes that hybrid technologies establish with both incumbent and emerging technologies. Data collection and simulation are conducted for investigating the relationship impacts and modes that HEVs as hybrid technologies establish with both ICEVs and BEVs as incumbent and emerging technologies, respectively. This is investigated under several scenarios including business-as-usual, landscape pressure, predator-prey, and sociotechnical transition over 1985-2070 in the US market.

The remainder of the paper is organised as follows. Section 2 presents the technology interaction framework. Section 3 presents the methodology including case study, and modelling and simulation methods. Results and findings are presented in section 4, followed by discussion, and concluding remarks, in section 5.

## 2. The technology interaction framework

The technology interaction framework proposed by (Mirzadeh Phirouzabadi, 2021; Mirzadeh Phirouzabadi et al., 2022) was adopted to conceptualise inter-technology relationships. The framework is based on the technological innovation system (TIS), defined as a network of actors that interact in a specific economic/industrial area and under a particular institutional setup to develop, diffuse and deploy a specific technology in society (Bergek et al., 2008; Hekkert et al., 2007; Suurs, 2009). The technology interaction framework is comprised of a set of multi-modal and multi-dimensional building blocks. The TIS structural and dynamical elements constitute the multi-dimensional building block in the framework. The TIS dynamical elements refer to the seven TIS processes, i.e., entrepreneurial activities, knowledge development and diffusion, guidance of the search, market formation, resource mobilisation, and legitimisation, that lead to the build-up of four TIS structural elements, i.e., actors, institutions, networks, and resources (Bergek et al., 2008; Hekkert et al., 2007; Suurs, 2009). Here, a technology is defined as a TIS with seven internal dynamics that build up four internal structural elements.

The biological relationship modes constitute the multi-modal building block in the framework. It was conceptualised that the internal dynamics of one TIS can become coupled with the internal dynamics of other TISs via what was referred to as 'co-dynamics', such as knowledge development and diffusion co-dynamics, entrepreneurial co-dynamics, and resource mobilisation co-dynamics. Such co-dynamics can eventually lead to the build-up of shared structural elements, known as structural 'couplings' or 'overlaps', between interacting TISs, such as overlap actors, knowledge overlaps, and resource overlaps (Binz & Truffer, 2017; Mäkitie et al., 2018; Mirzadeh Phirouzabadi et al., 2020a, 2022; Mirzadeh Phirouzabadi et al., 2020b; Mirzadeh Phirouzabadi et al., 2020c; Mirzadeh Phirouzabadi et al., 2020g). For example, knowledge development co-dynamics are established between two interacting TISs when the internal knowledge development dynamics of the two TISs influence and couple with each other via processes such as knowledge recombination or transfer. These knowledge development co-dynamics can lead to knowledge overlaps between the interacting TISs (Mäkitie et al., 2018; Mirzadeh Phirouzabadi et al., 2020c). Co-dynamics trade externalities between the interacting TISs, which can be featured with directionality (bi- or unilateral), impact (positive, negative or neutral), and intensity (weak or strong) (Bergek & Onufrey, 2013; Markard & Hoffmann, 2016; Mirzadeh Phirouzabadi et al., 2022; Mirzadeh Phirouzabadi et al., 2020g; Onufrey & Bergek, 2015). A combination of positive, negative, or neutral externalities between the interacting TISs lead to the establishment of one of the six biological relationship modes between them, i.e., competition, symbiosis, parasitism, commensalism, amensalism, and neutralism (Mirzadeh Phirouzabadi et al., 2020g; Pistorius & Utterback, 1997; Sandén & Hillman, 2011). For instance, a symbiotic mode can be observed through knowledge development co-dynamics of both HEVs and BEVs as both can benefit from the technological knowledge advancements in their shared components such as batteries, electric engines and engine control systems (Dijk, 2014). Or an amensalism mode can be observed through entrepreneurial co-dynamics of both ICEVs and BEVs when General Motors (GM), as the EV flagship auto manufacturer, destabilised the further development and mass commercialisation of its EV1 in the early 2000's because the market was found to be a huge threat to the ICEVs value chain (Mirzadeh Phirouzabadi et al., 2020a).

## 3. Methodology

### 3.1. The case of ICEVs, HEVs and BEVs in the US market

The case of ICEVs, HEVs and BEVs in the US market suits this research, firstly, because it accommodates suitable candidates for incumbent, hybrid and emerging technologies in an industry. While ICEV is known as an incumbent technology with competence-sustaining innovations, BEV is known as a disrupting technology with competence-destroying innovations (Bergek et al., 2013). BEV replaces the entire components of ICEV with several components including charger, power convertor and controller, battery packs and electric or traction motor (Poullikkas, 2015). HEV is known as a hybrid technology with competence-expanding innovations as it combines the components of both ICEV and HEV, but in a smaller version (Bergek et al., 2013; Oltra & Saint Jean, 2009). Secondly, the hybridisation dilemma persists in the case of HEV, as pointed out by Sperling and Lipman (2003), "Are hybrid vehicles likely to dominate? [...] Are they a second-best option that will be delayed in the near term and succumb to fuel cells and other technologies in the long term?". Thirdly, these powertrain systems are characterized by systemic properties and can be entrenched into a wide range of aspects such as user preferences, cultural beliefs, technical knowledge, company capabilities, infrastructure and public authorities (Bergek et al., 2013; Dijk, 2014; Dijk et al., 2015). Finally, the required data and information to conduct our research were available and well documented for the case of powertrain systems in the US market (Mirzadeh Phirouzabadi, 2021; Mirzadeh Phirouzabadi et al., 2022; Mirzadeh Phirouzabadi et al., 2020d; Mirzadeh Phirouzabadi et al., 2020f).

### 3.2.System Dynamics (SD) Modelling

The technology interaction framework was constructed and simulated in an extensive System Dynamics (SD) modelling since SD models are based upon virtuous and vicious loops, cumulative causations, time delays, and complex, non-linear relationships, which suit the nature of technological evolution and interactions (Raven & Walrave, 2020; Sterman, 2000; Walrave & Raven, 2016). While the full description of the SD model is available in Mirzadeh Phirouzabadi et al. (2022), it is briefly described in the following.

The dynamics and co-dynamics of interacting TISs are quantified using the Lotka-Volterra (L-V) equations. They were originally developed and applied for inter-population relationships in biological ecosystems (Lotka, 1926; Volterra, 1927), but have also been applied to inter-technology relationships (Kim et al., 2006; Kreng et al., 2012; Marasco et al., 2016; Mirzadeh Phirouzabadi et al., 2020a; Mirzadeh Phirouzabadi et al., 2020b; Mirzadeh Phirouzabadi et al., 2020c; Pretorius et al., 2015). The L-V equations can accommodate a variety of mathematical models including simple or decaying exponential functions, logistic model (Song & Aaldering, 2019), Bass model (McManus & Senter Jr, 2009) and Gompertz model (Muraleedharakurup et al., 2010). They are comprised of the parameters of $a$ and $b$, and $c$. While the first two parameters represent the growth or decline rate of dynamics, the last parameter represents the growth or decline rate of co-dynamics.

(i) The parameter $a$ represents the growth rate of a TIS dynamic when the TIS develops in isolation from other TISs in the market. For instance, the more positive the value of $a$ for a TIS knowledge dynamic the more the TIS favours knowledge variety and creation (Table 1). However, the more negative the value the more the TIS lacks creativity and opposes knowledge variety and creation (Castiaux, 2007; Mirzadeh Phirouzabadi et al., 2020c).

(ii) The parameter $b$ represents the decline rate of a TIS dynamic, again when the TIS develops in isolation in the market. For instance, the more positive the value of $b$ for for a TIS knowledge dynamic, the more exploitative the TIS behaves, which means it tends to archive, control, maintenance and exploit its knowledge (Table 1). The more negative the value, the more explorative the TIS behaves, which means it tends to value emergent matters, flexibility and knowledge exploration in order to create radical innovations (Castiaux, 2007; Mirzadeh Phirouzabadi et al., 2020c).

(iii) The parameter $c$ as the external interaction rate represents the growth or decline rate of co-dynamics of interacting TISs. It refers to the rate at which the growth of a TIS dynamic is affected by the presence of the same dynamic of other TISs. The relationship modes between interacting TIS are determined by estimating the sign of their external interaction rates. Table 2 exemplifies inter-powertrain relationship signs and modes for the case of HEV and BEV.

Table 1- interpretation for the negative and positive sign of growth and decline rates of HEV (Mirzadeh Phirouzabadi et al., 2020c)

| Parameters | + | - |
|---|---|---|
| $a_{HEV}$ growth rate | creative | uncreative |
| $b_{HEV}$ decline rate | exploitative | explorative |

Table 2- The L-V signs of relationship modes between HEV and BEV, adopted from Mirzadeh Phirouzabadi et al. (2020c) and Pistorius and Utterback (1997)

| Mode of interaction | $c_{BEV-HEV}$ | $c_{HEV-BEV}$ |
|---|---|---|
| Competition | + | + |
| Symbiosis | - | - |
| Neutralism | 0 | 0 |
| Parasitism | -(+) | +(-) |
| Commensalism | 0(-) | -(0) |
| Amensalism | 0(+) | +(0) |

### 3.3. Operationalisation

The SD technology interaction model was operationalised and simulated in the VENSIM environment (DSS 8.0.9). Taking the unit of time as years, the time horizon of the simulation is 1985 to 2070. Euler's algorithm method was used with a step size (dt) of 0.125 of a year. The SD model was validated through the VENSIM built-in capabilities, including calibration, extreme case testing, sensitivity analysis and reality checks (Mirzadeh Phirouzabadi et al., 2022; Raven & Walrave, 2020; Walrave & Raven, 2016). The detailed operationalisation including stocks, flows, and constants, and the validation results are presented in Mirzadeh Phirouzabadi et al. (2022).

The database provided by Mirzadeh Phirouzabadi et al. (2022) and Mirzadeh Phirouzabadi (2021) were used to run the simulation. These data were collected from various sources including Scopus, Thomson Reuters, Google Trends, the US Alternative Fuels Data centre, the US FuelEconomy, the US Bureau of Economic Analysis, and the US Environmental Protection Agency. The main indicators and measurements are briefly described here.

- The knowledge development dimension was determined by two sub-dimensions: (i) scientific knowledge development, measured by number of publications, and (ii) technological knowledge development, measured by number of patents.
- The knowledge diffusion dimension was determined by two sub-dimensions: (i) scientific and technological forward citations, measured by number of publication forward citations and number of patent forward citations, respectively, and (ii) scientific and technological inter-organisational collaborations, measured by number of bilateral relationships in joint publications and number of bilateral relationships in joint patents, respectively.

- The entrepreneurship activities dimension was determined by two sub-dimensions: (i) scientific and technological entrepreneurships, measured by number of publication assignees and number of patent assignees, respectively, and (ii) vehicle models, measured by number of vehicle models introduced to the market.
- The guidance of the search dimension was determined by two sub-dimensions: (i) regulative performance, measured by number of laws and regulations, and (ii) search popularity, measured by Google search popularity.
- The market formation dimension was determined by two indicators: (i) incentive performance, measured by number of incentives, and (ii) market share, measured by share of vehicle sales.
- The resource mobilisation dimension was determined by two sub-dimensions: (i) scientific and technological manpower, measured by number of publications authors and number of patents applicants, respectively, and (ii) financial mobilisation, measured by amount of financial capital which included several factors such as the monetary value of publications and patents, the fixed and operating ownership costs (e.g., manufacturer suggested retail price, fuel price (e.g., electricity and gas prices), maintenance cost (e.g., refuelling infrastructure installation cost, battery replacement cost, tyre cost, and routine maintenance costs)), and well-to-wheel (WTW) environmental and energy costs.

To gain an overall understanding of observed patterns, all the dimensions were grouped into two categories of technology development side (comprised of dimensions related to the technology development of a TIS such as scientific and technological knowledge development and citations, and scientific and technological manpower) and the market development side of a TIS (comprised of dimensions related to market development such as entrepreneurship, Google popularity, incentives, resource mobilisation, and market share). For explorative-exploitative analysis, for example, we summed the value of the growth rate of (sub) dimensions.

### 3.4.Scenarios

Seven scenarios were constructed to investigate whether and under what circumstances HEV would accelerate or suspend the transition from ICEV to BEV (Mirzadeh Phirouzabadi, 2021; Mirzadeh Phirouzabadi et al., 2022):

1. Baseline scenario: This scenario presents the business-as-usual (BAU) situation, as it reflects the historical trend and current projections of chosen variables. All the three incumbent, hybrid and emerging powertrain systems are present and interact in the market.
2. Landscape pressure scenario: This scenario created an extremely unfavourable environment for ICEV by setting higher oil price (150% BAU), higher tax and registration fees (150% BAU), higher gross domestic product (GDP) growth rate (150% BAU), and higher WTW costs (150% BAU) (Mirzadeh Phirouzabadi et al., 2022; Pasaoglu et al., 2016).
3. Niche-incumbent scenario: This scenario eliminated HEV from the market; hence, only ICEV and BEV were present.
4. Hybrid-incumbent scenario: This scenario eliminated BEV from the market; hence, only HEV and ICEV were present.
5. Sociotechnical transition scenario: In this scenario when BEV exhibited a structural decline[1] its dynamics were reinforced for 20 years but at the same time the dynamics of both ICEV and HEV were weakened (only if HEV market share was reached beyond 50%).
6. Niche-favoured relationships scenario: In this scenario, all the co-dynamics (or inter-powertrain relationship modes) observed in baseline scenario were either strengthened or weakened in favour of BEV. For instance, a symbiotic relationship mode observed between HEV and BEV in baseline scenario would be kept unchanged, but its size would be altered in a way that more positive externalities could be provided from HEV towards BEV, and less positive externality the other way around.
7. Predator-prey scenario: This scenario presents the primary predator-prey relationship in the transitionary situation where BEV and ICEV established a pure competitive relationship, HEV and ICEV established a parasitic relationship (the latter as prey), and HEV and BEV established a parasitic relationship (the latter as predator). Hence, any relationship modes observed in baseline scenario would be changed, accordingly.

[1] It was assumed that BEV would exhibit a structural decline when there was a sustained negative trend in its market growth, i.e., the value of the 3-year moving average of 'change in its market share' was found negative (Raven & Walrave, 2020).

## 4. Results

HEV projected a creative and explorative behaviour on the technology development side in baseline scenario (Figure 1a). This was also observed on the individual technology development dimensions, such as scientific and technological knowledge development (Figure 2a and Figure 2b, respectively). This, for instance, means that HEV mainly favoured scientific and technological knowledge variety and creation and it mostly explored scientific and technological knowledge throughout the period. On the contrary, a more exploitative behaviour was observed on the market development side (Figure 1a) and the individual market development dimensions, such as Google search popularity (Figure 2c) and market share (Figure 2d). For market share, for instance, this means that HEV rather favoured controlling and exploiting the existing market throughout the period.

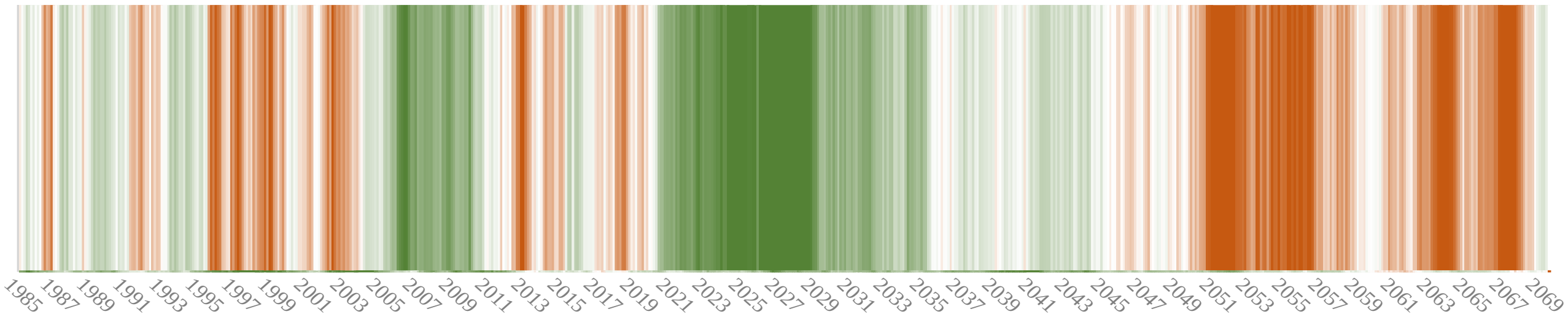

(a) Technology development dimension

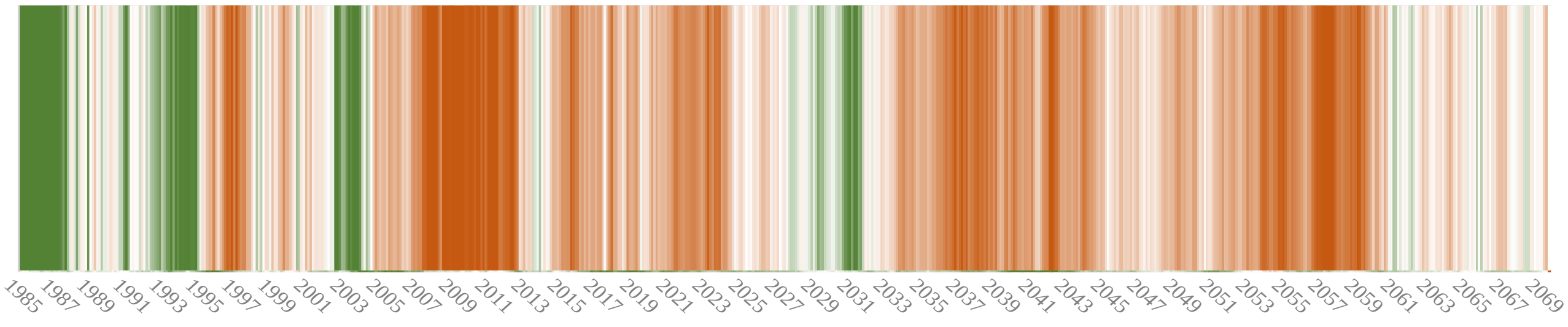

(b) Market development dimension

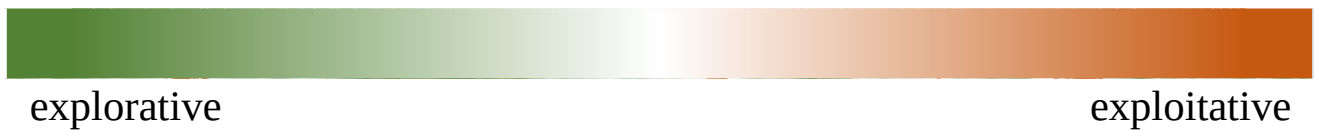

Figure 1. HEV's explorative-exploitative behaviour on technology and market development sides (baseline scenario)

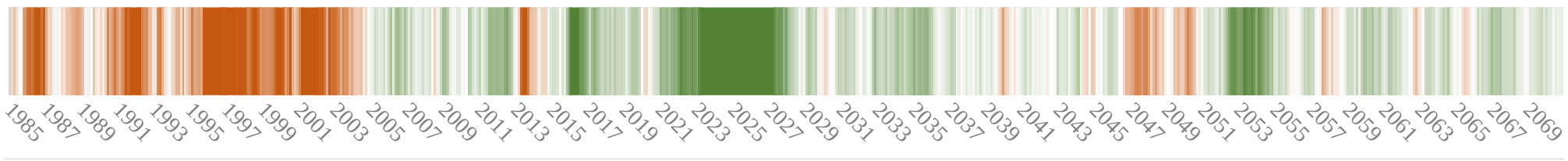

(a) Scientific knowledge development

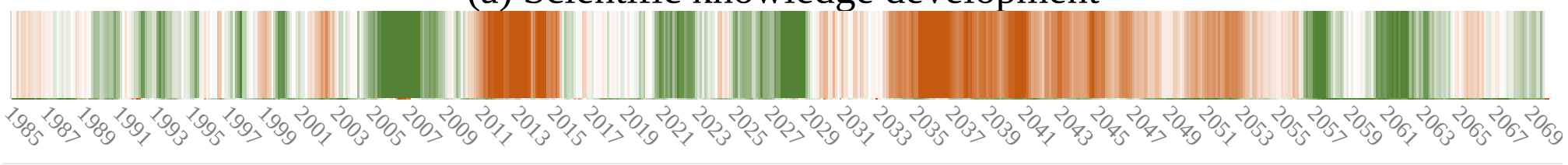

(b) Technological knowledge development

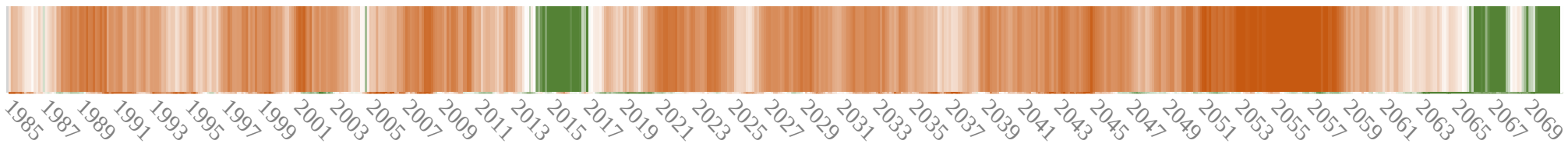

(c) Google Search popularity

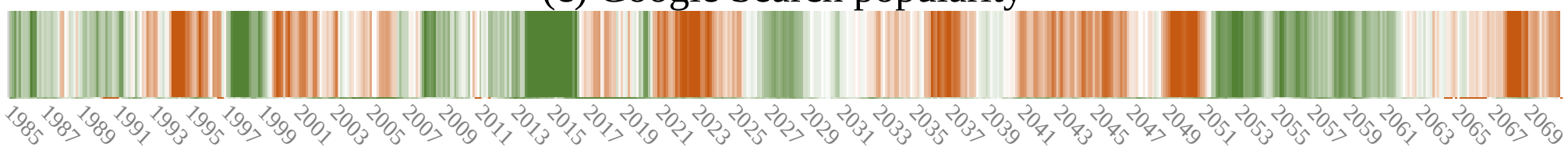

(d) Market share

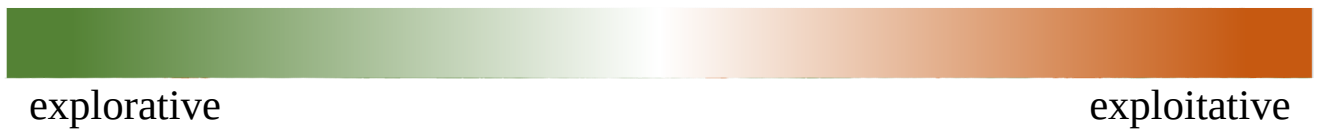

Figure 2. HEV's explorative-exploitative behaviour on individual dimensions (baseline scenario)

Regarding relationship modes, HEV mainly established a mixed mode of symbiotic, parasitic, and commensal relationships with both ICEV and BEV for the dimensions scientific and technological knowledge development, technological organisations, and market share (Figure 3 and Figure 4, respectively). For instance, BEV benefited from the scientific knowledge development of HEV via the modes of symbiosis, parasitism and commensalism in both baseline and landscape pressure scenarios (Figure 3a). However, Figure 5 depicts that HEV mainly established a symbiotic relationship mode with the other two powertrain systems on the technology development side, where HEV happened to provide more positive externalities for the technology development side of BEV than for ICEV. The more positive externalities could be a reason for observing a significant decline on BEV's technological development side in the absence of HEV, i.e., niche-incumbent scenario (Figure 6). The situation is different on the market development side as the results revealed a mixed mode of relationships for HEV-BEV (Figure 7a) and HEV-ICEV (Figure 7b). On the one side, BEV supported and reinforced the market development side of HEV via symbiotic, parasitic, and commensal relationships, except between the late 2020s and the early 2030s, and the late 2040s (Figure 7a). On the other side, HEV either inhibited or did not support the market development of BEV via parasitic and commensal relationships, except in the late 1990s and in the mid-2020s (Figure 7a). This justifies why BEV's market share increased substantially in the absence of HEV (i.e., hybrid-incumbent scenario), especially after the mid-2030s (Figure 6e). Both HEV and ICEV established a symbiotic relationship on the market development side for a long period of time, though HEV only shortly (between the late 2020s and the early 2030s) acted as a predator towards ICEV's market development side (Figure 7b).

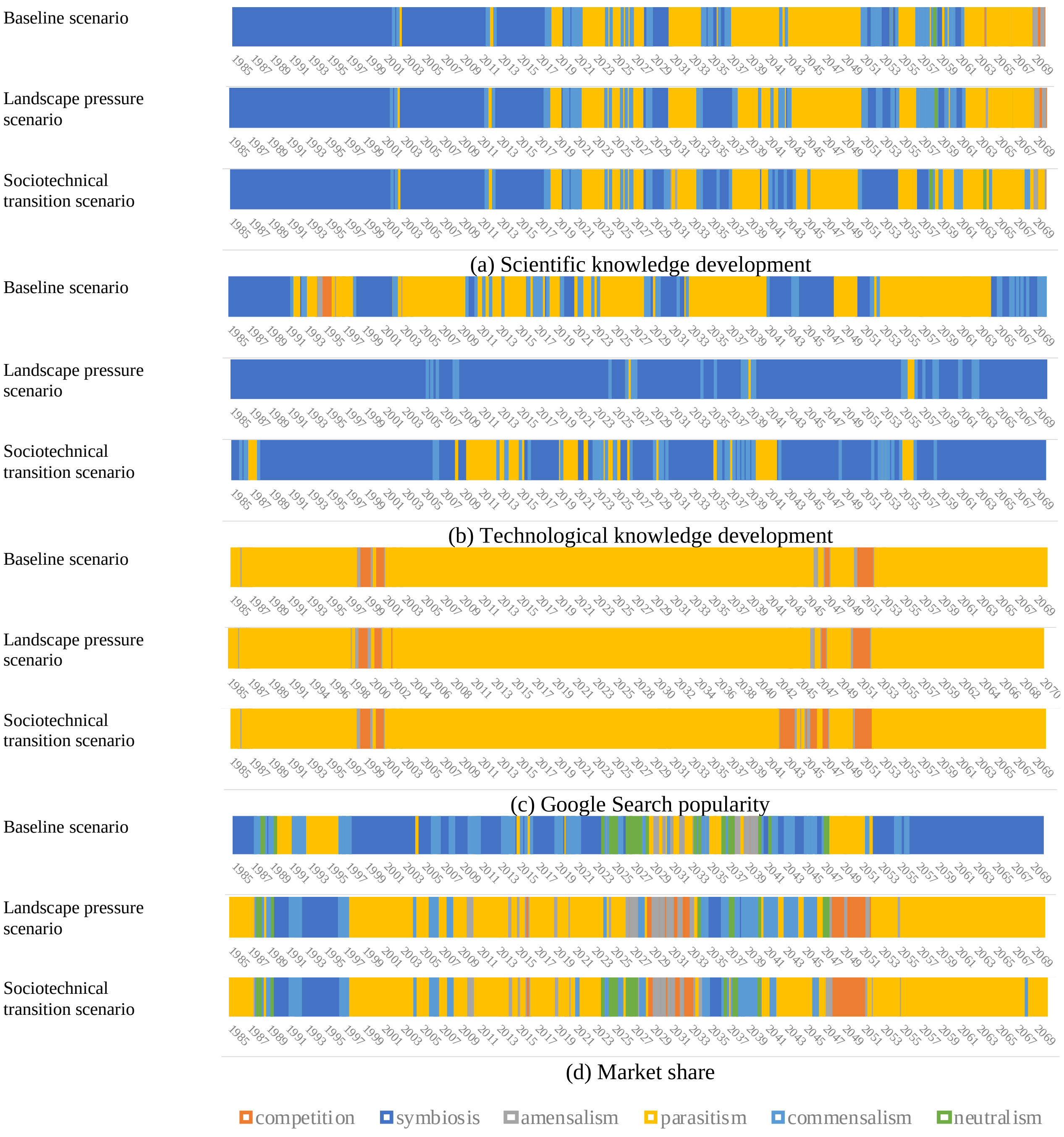

Figure 3. BEV–HEV relationship modes

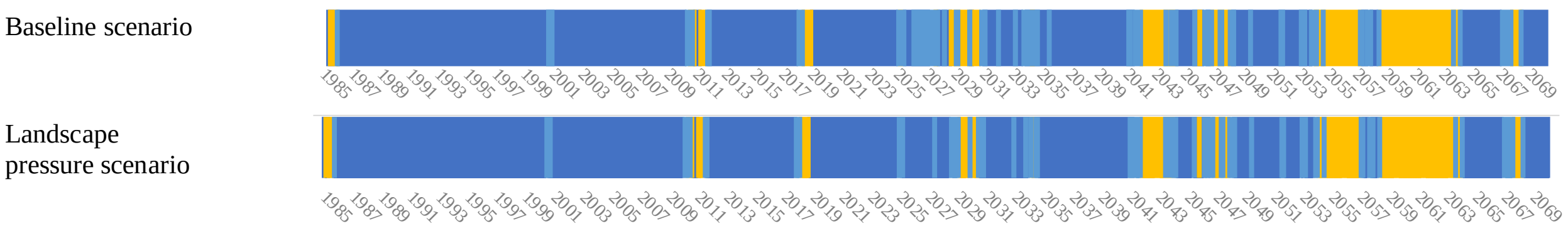

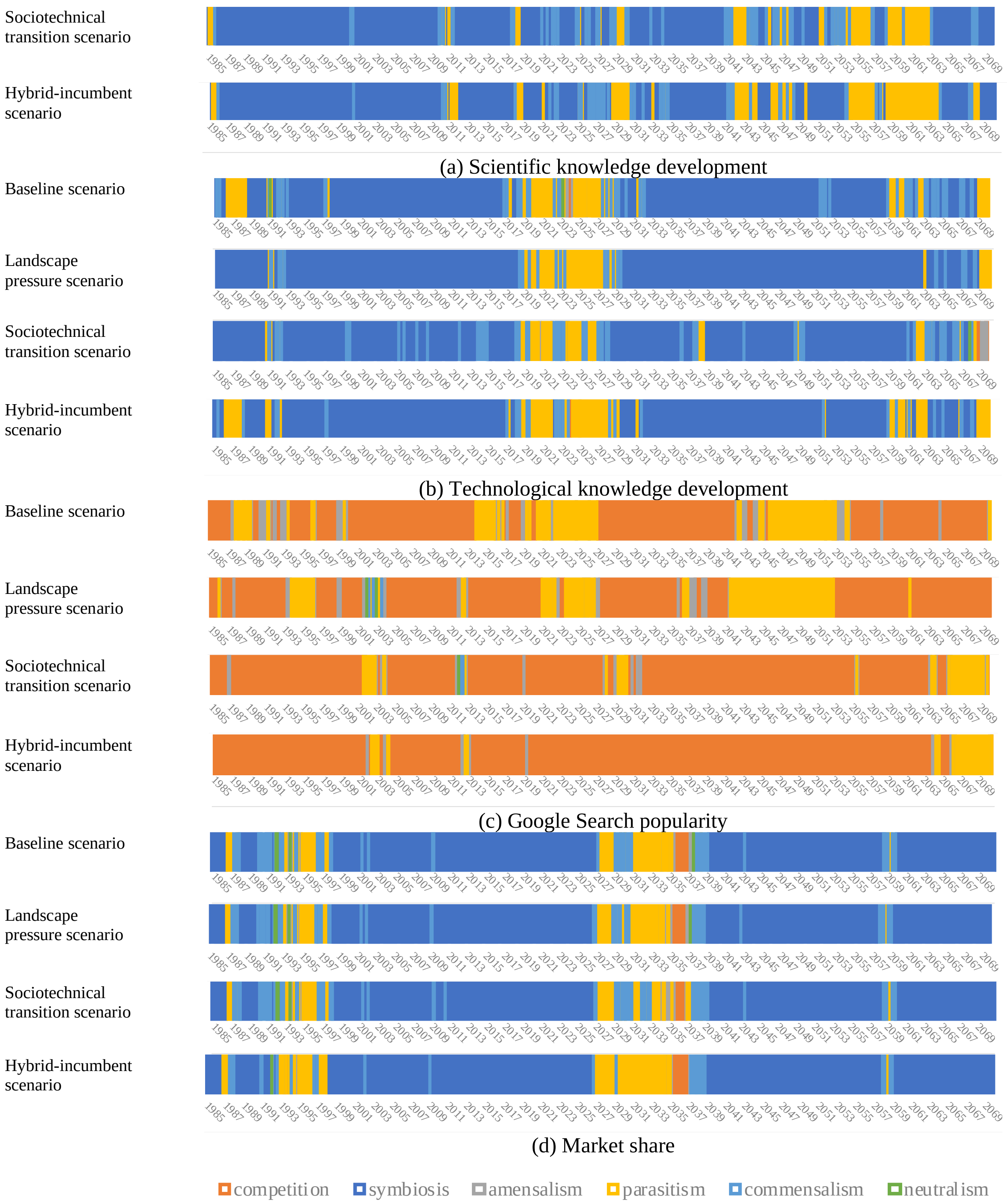

Figure 4. HEV–ICEV relationship modes

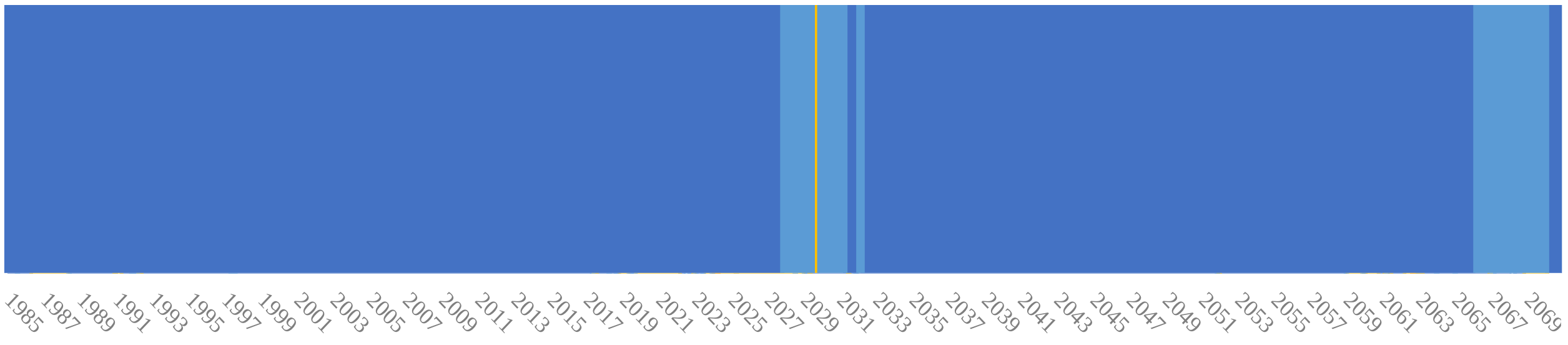

(a) BEV–HEV

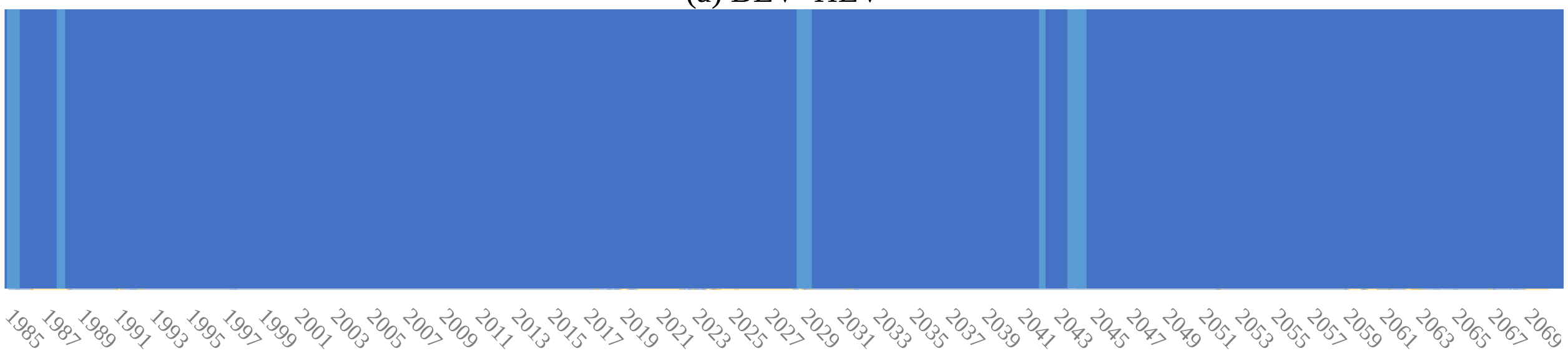

(b) ICEV–HEV

competition symbiosis amensalism parasitism commensalism neutralism

Figure 5. Aggregated relationship modes on technology development side (baseline scenario)

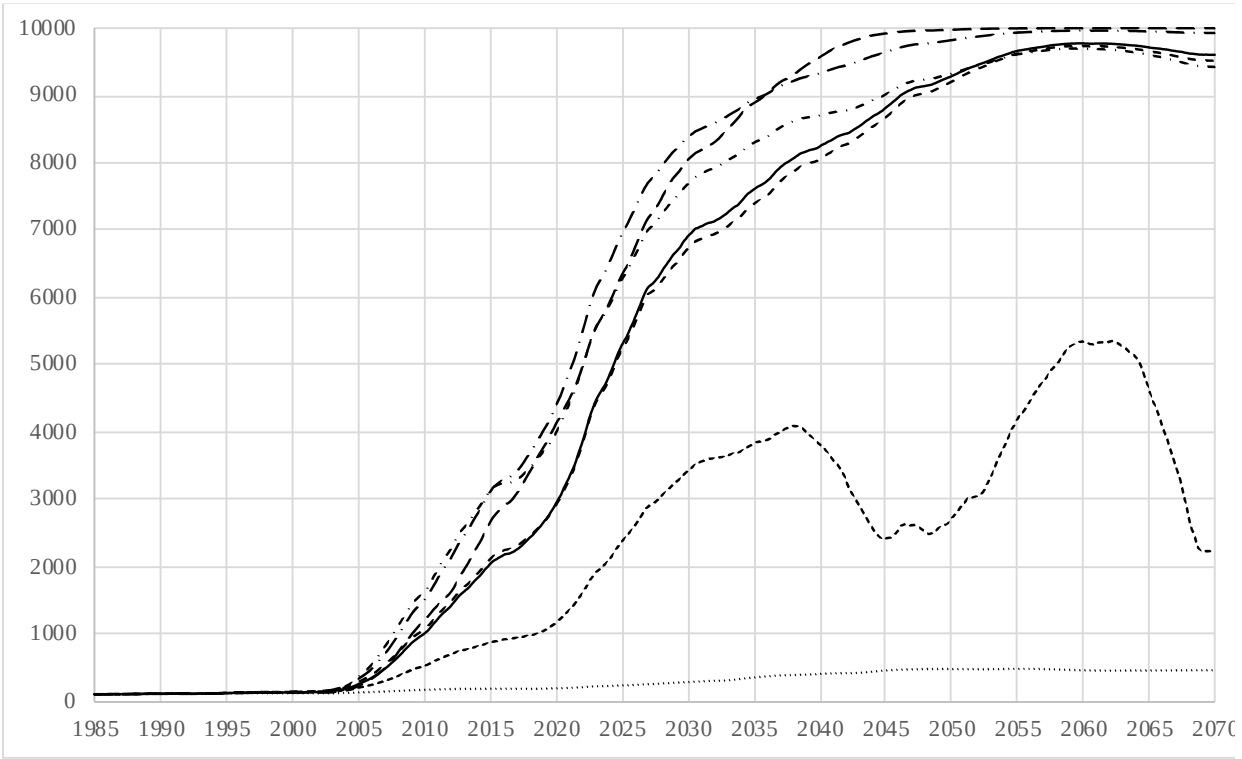

(a) Number of publications

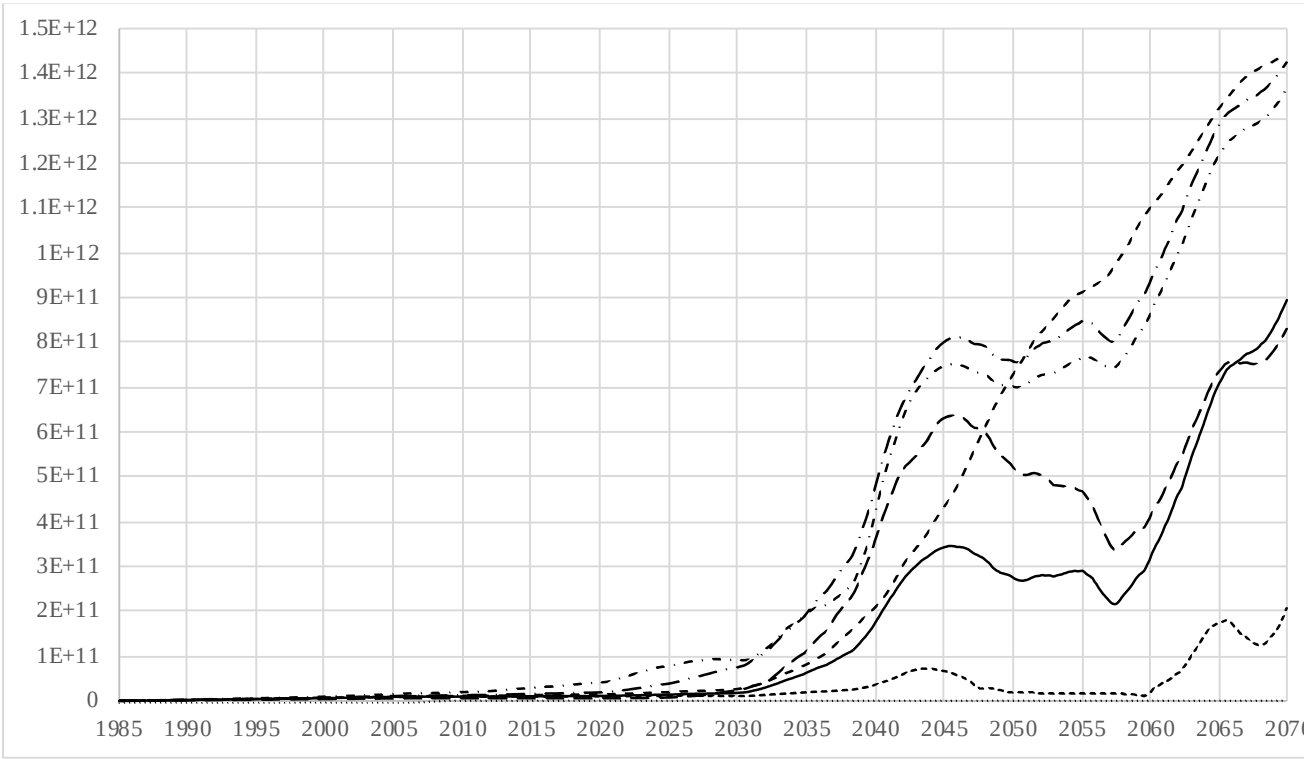

(c) Amount of financial mobilisation

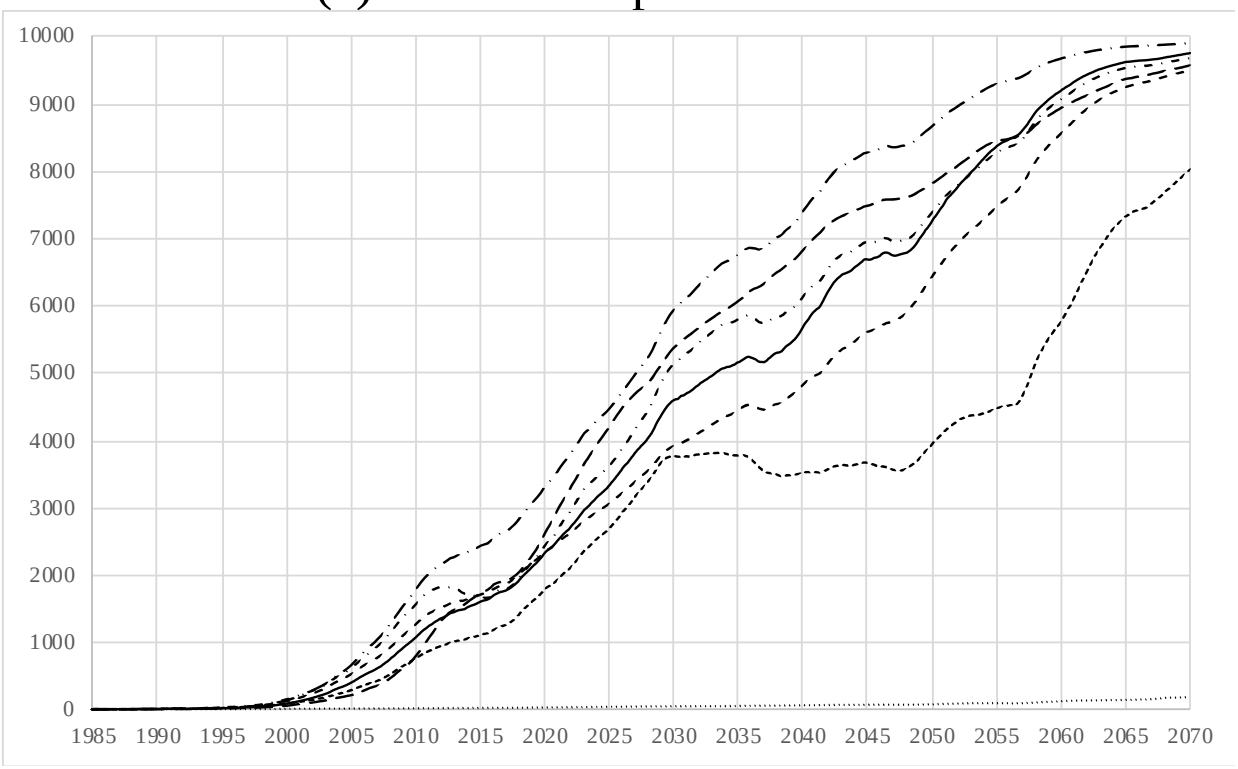

(b) Number of patents

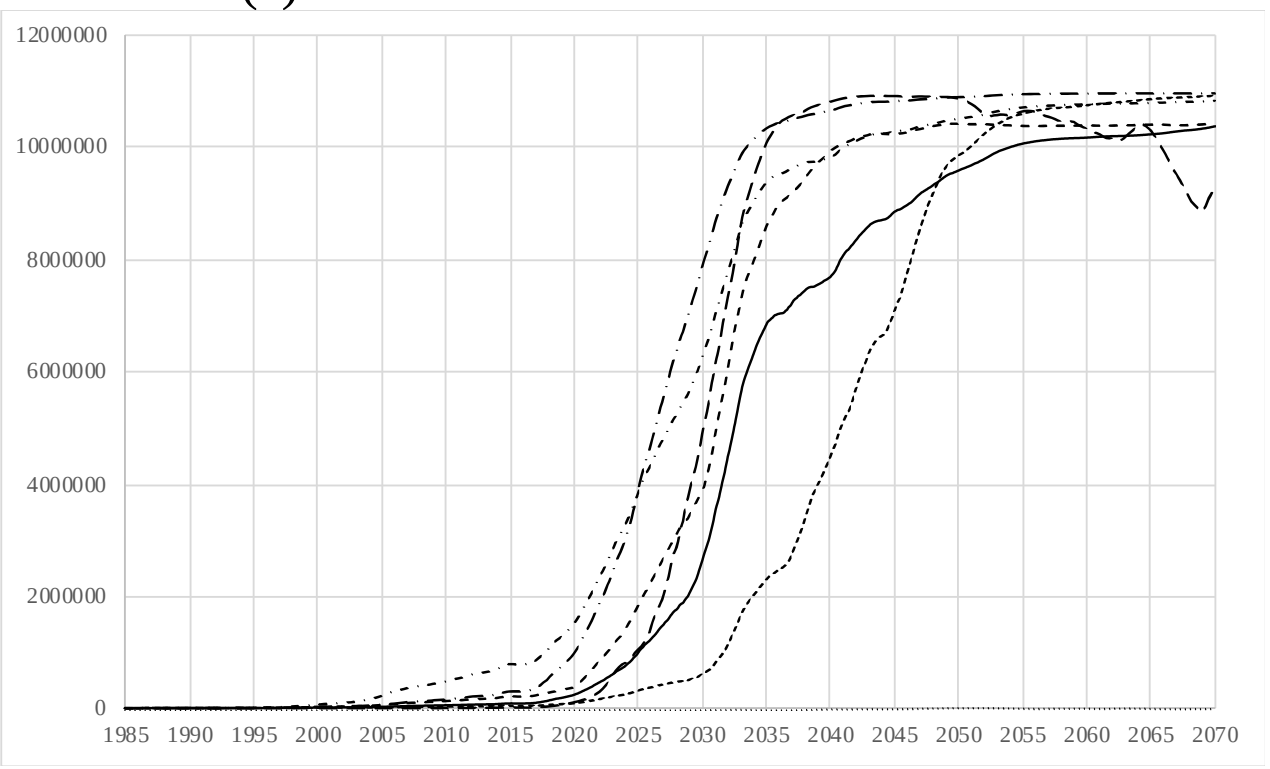

(d) Number of vehicle sales

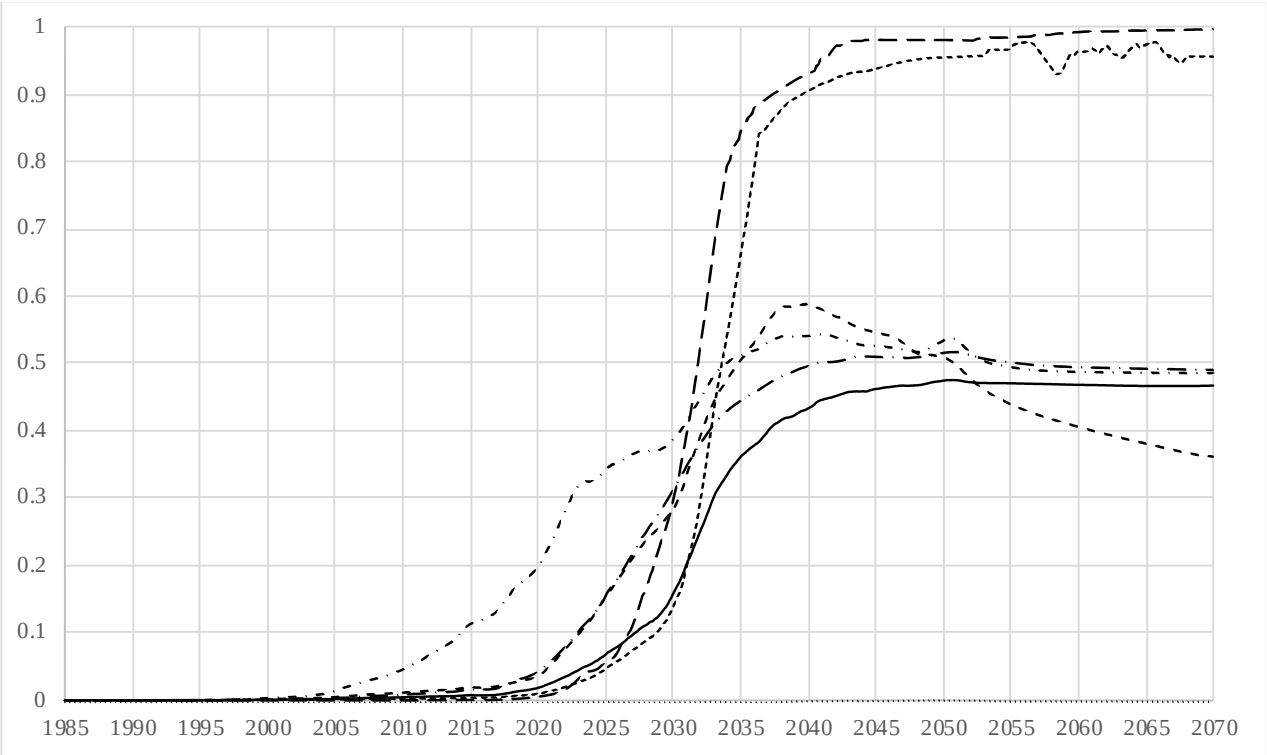

(e) Market share

Baseline | Hybrid-incumbent | Niche-incumbent | Landscape press
Sociotechnical transition | Niche favoured relationships | Predator-prey

Figure 6. BEV's (sub)dimensions under seven scenarios

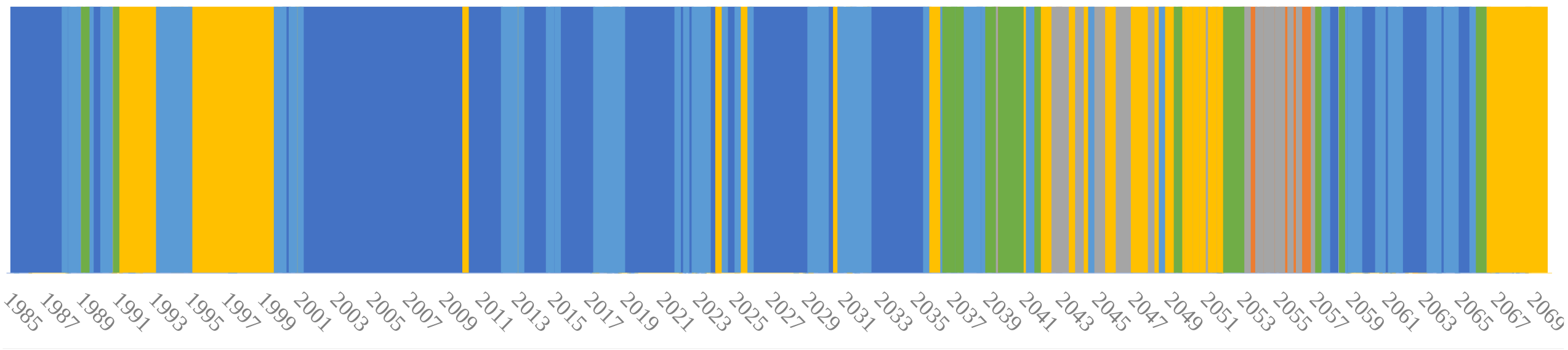

(a) BEV–HEV

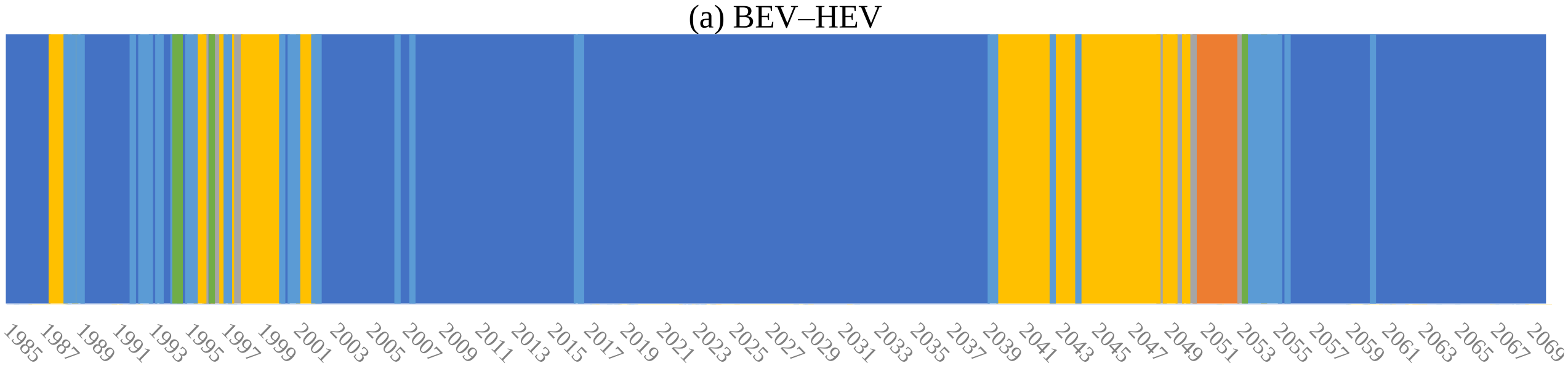

(b) ICEV–HEV

competition symbiosis amensalism parasitism commensalism neutralism

Figure 7. Aggregated relationship mode on market development side (baseline scenario)

When HEV was not present in the industry (i.e., niche-incumbent scenario), BEV experienced a significant decline in most dimensions. For instance, a significant decline was observed in the dimensions scientific and technological knowledge developments and resource mobilisation throughout the entire period (Figure 6a, Figure 6b, and Figure 6c, respectively). BEV's sales grew slower until the

mid-2030's in the absence of HEV[2], but they increased so dramatically after the mid-2040s that the market experienced 2 million additional BEV sales by the end of the period (roughly 10 million (95%) by 2050, as shown in Figure 6d and Figure 6e). While ICEV experienced an increase in some of its dimensions under landscape pressure scenario, such as scientific knowledge development and vehicle models (Figure 8a, Figure 8b, and Figure 8c, respectively), it possessed the lowest level of scientific knowledge development under hybrid-incumbent scenario, i.e., in the absence of BEV (Figure 8a). Additionally, the level of some of ICEV's dimensions, such as vehicle sales and share (Figure 8d and Figure 8e, respectively), declined much slower in landscape pressure scenario than in baseline scenario. Its market share in the absence of HEV (i.e., niche-incumbent scenario) started shrinking much later as it reached below 50% only after the mid-2030s.

GHG emissions results (Figure 9) revealed that the industry reached the highest accumulative GHG emissions level under hybrid-incumbent scenario. But the lowest accumulative GHG emissions level was observed under sociotechnical transition scenarios. This is because while BEV's market sales took off as early as the 1990's ICEV's market sales dropped so drastically in sociotechnical transition scenarios (compared to the 2000's in the other scenarios, Figure 8d). It is noticeable that even though favouring BEV's relationships (as in niche-favoured relationship scenario) led to an increase in the sales of BEVs in the market, it did not lead to lower accumulative GHG emissions level, compared to baseline scenario. This also should be noted that while HEV was forced to play a transitionary role in the market in predator-prey scenario, which led to a higher market share and sale for BEV (Figure 6d, and Figure 6e) and at the same time lower market share and sales for ICEV (Figure 8d, and Figure 8e, respectively), it still resulted in a higher amount of GHG emissions, compared with socio-technical scenario.

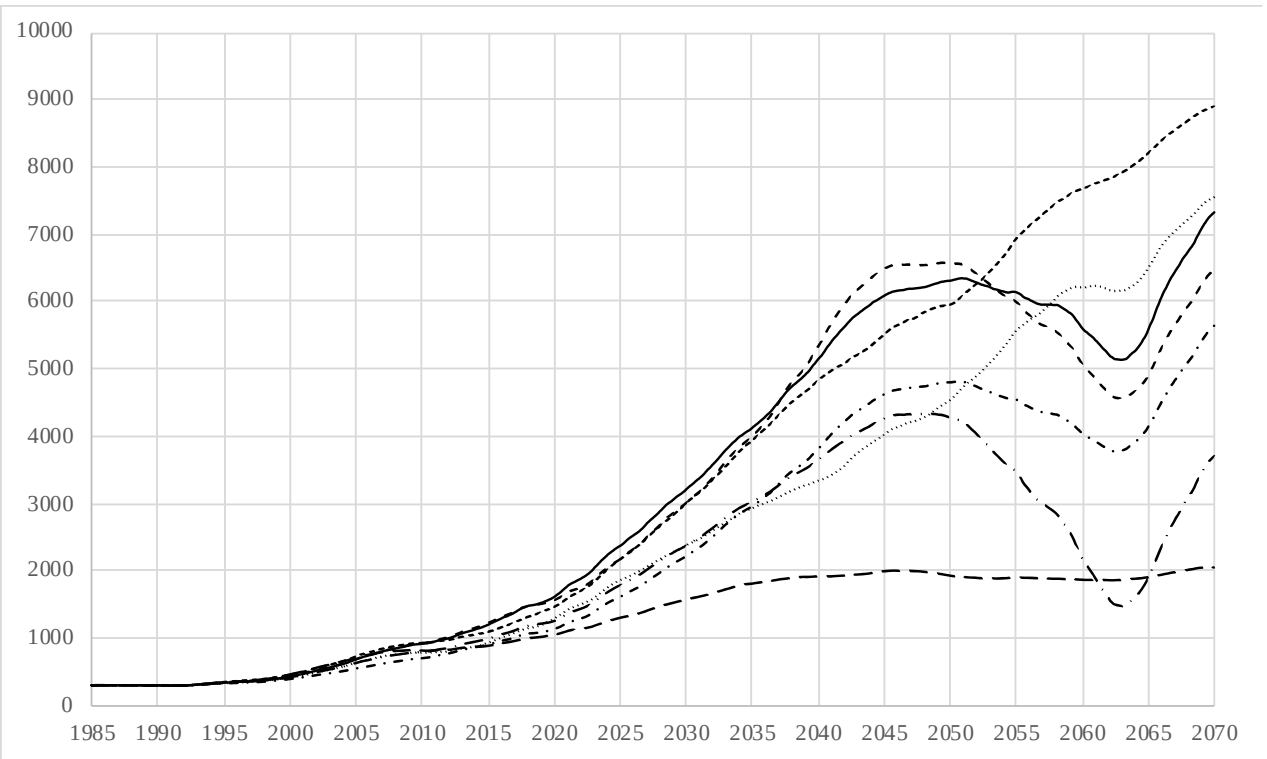

(a) Number of publications

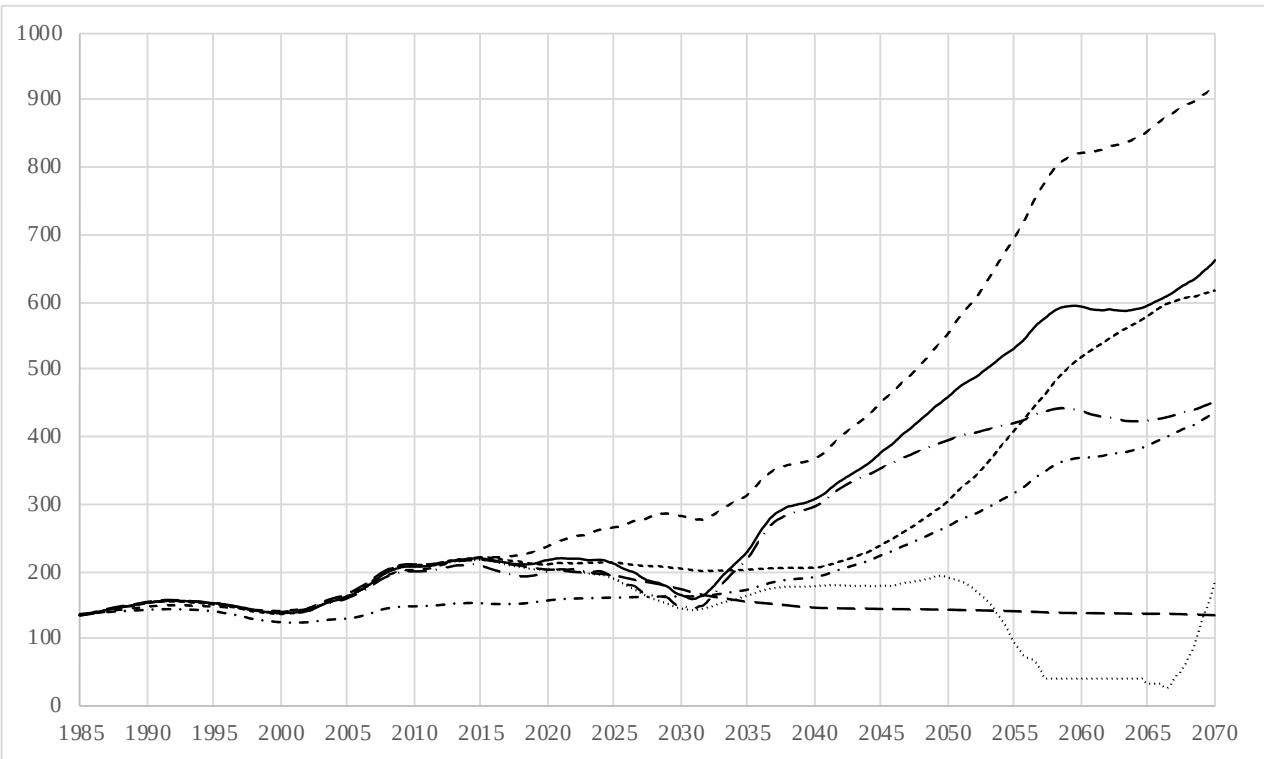

(c) Number of vehicle models

[2] For instance, around 500,000 and 4.5 million sales in 2030 and 2040, respectively, were observed in niche-incumbent scenario compared with over 2 and 6 million sales in the respective years in baseline scenario (Figure 6f).

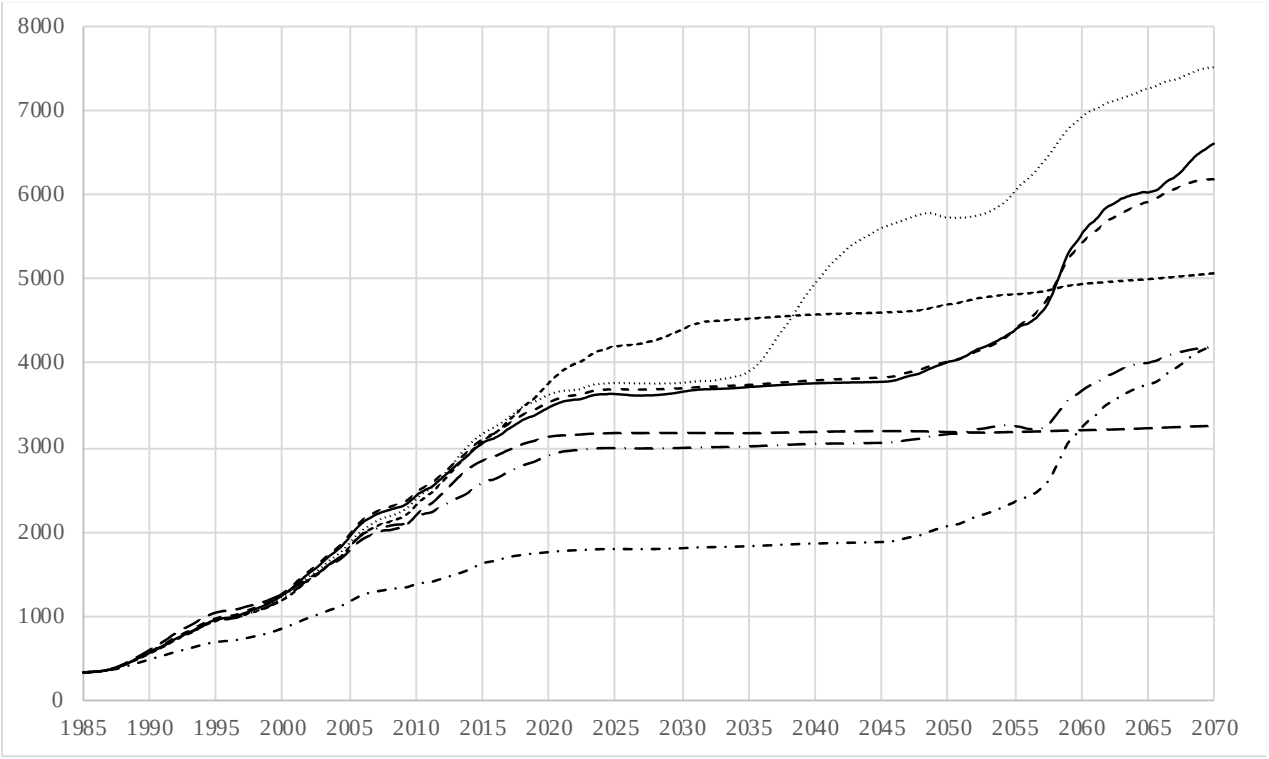

(b) Number of patents

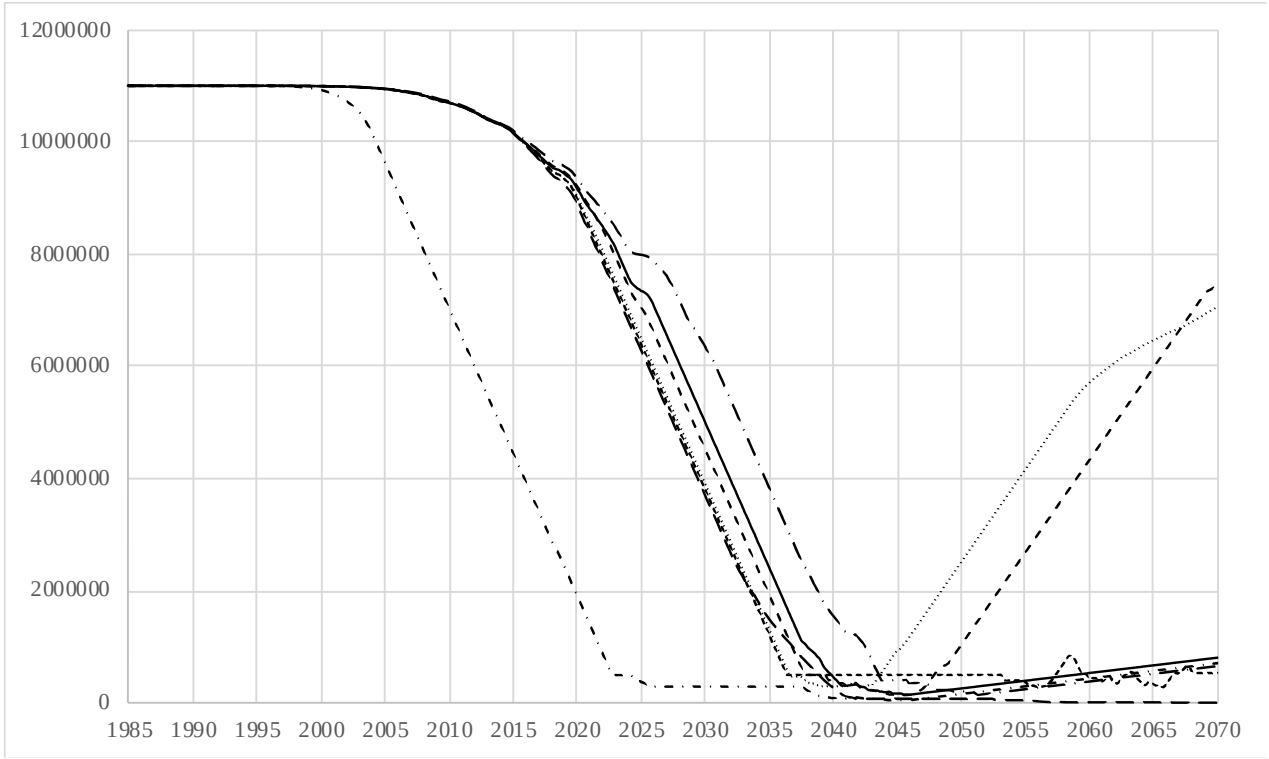

(d) Number of vehicle sales

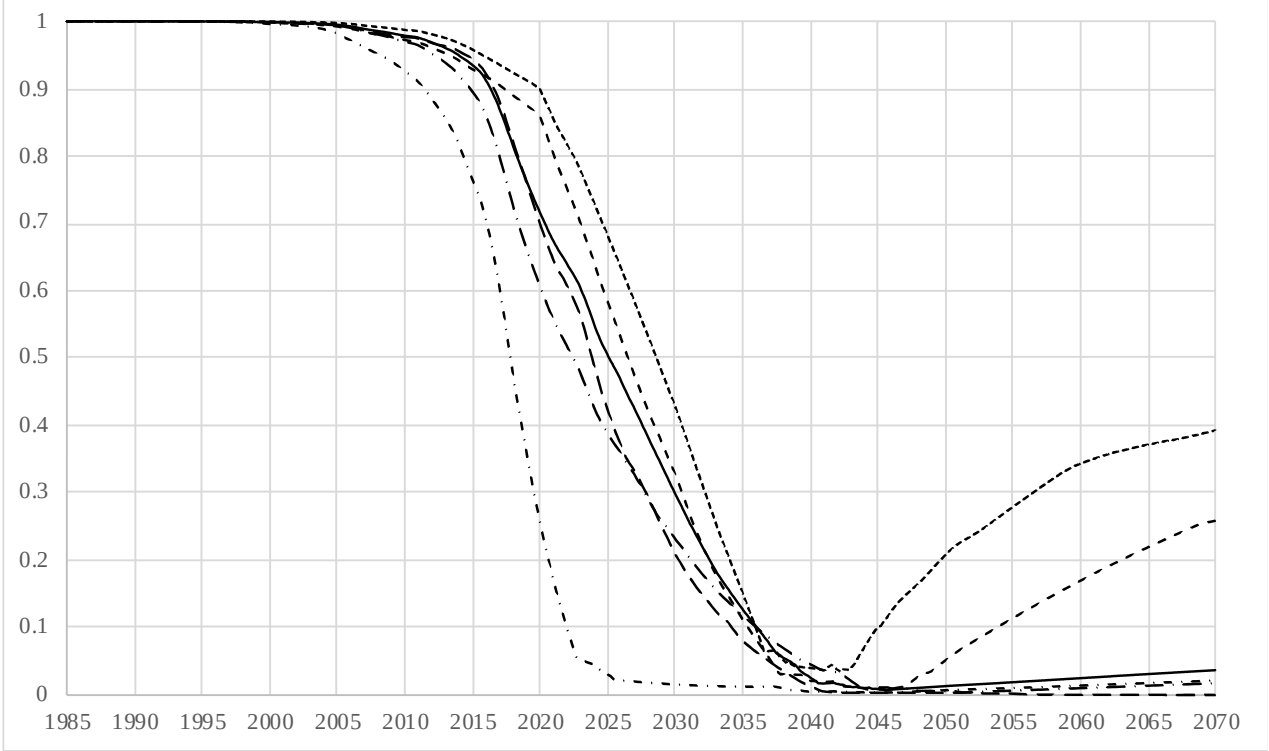

(e) Market share

Baseline | Hybrid-incumbent | Niche-incumbent | Landscape pressure
Sociotechnical transition | Niche favoured relationships | Predator-prey

Figure 8. ICEV’s (sub)dimensions under seven scenarios

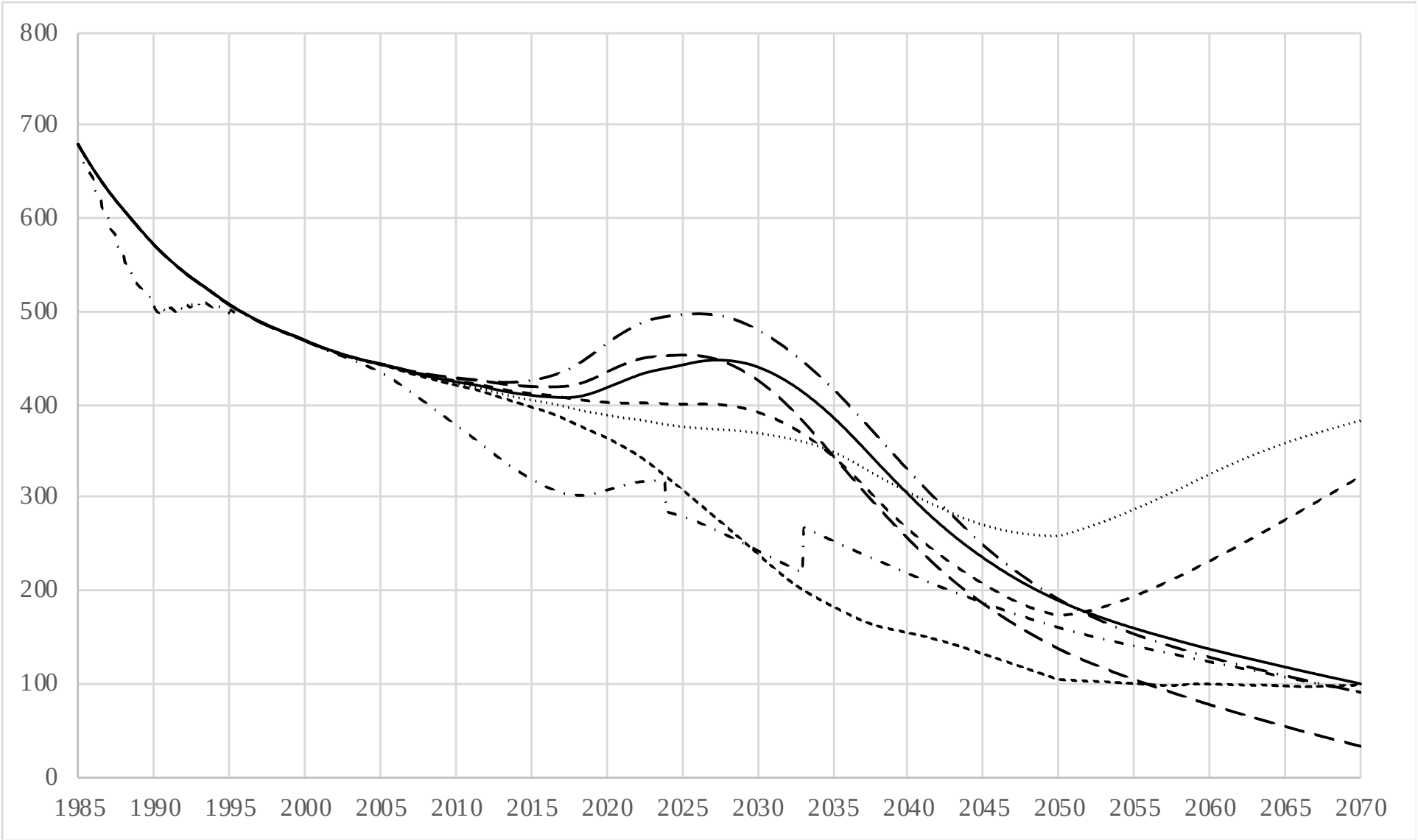

(a) Absolute GHG emissions (Mtons)

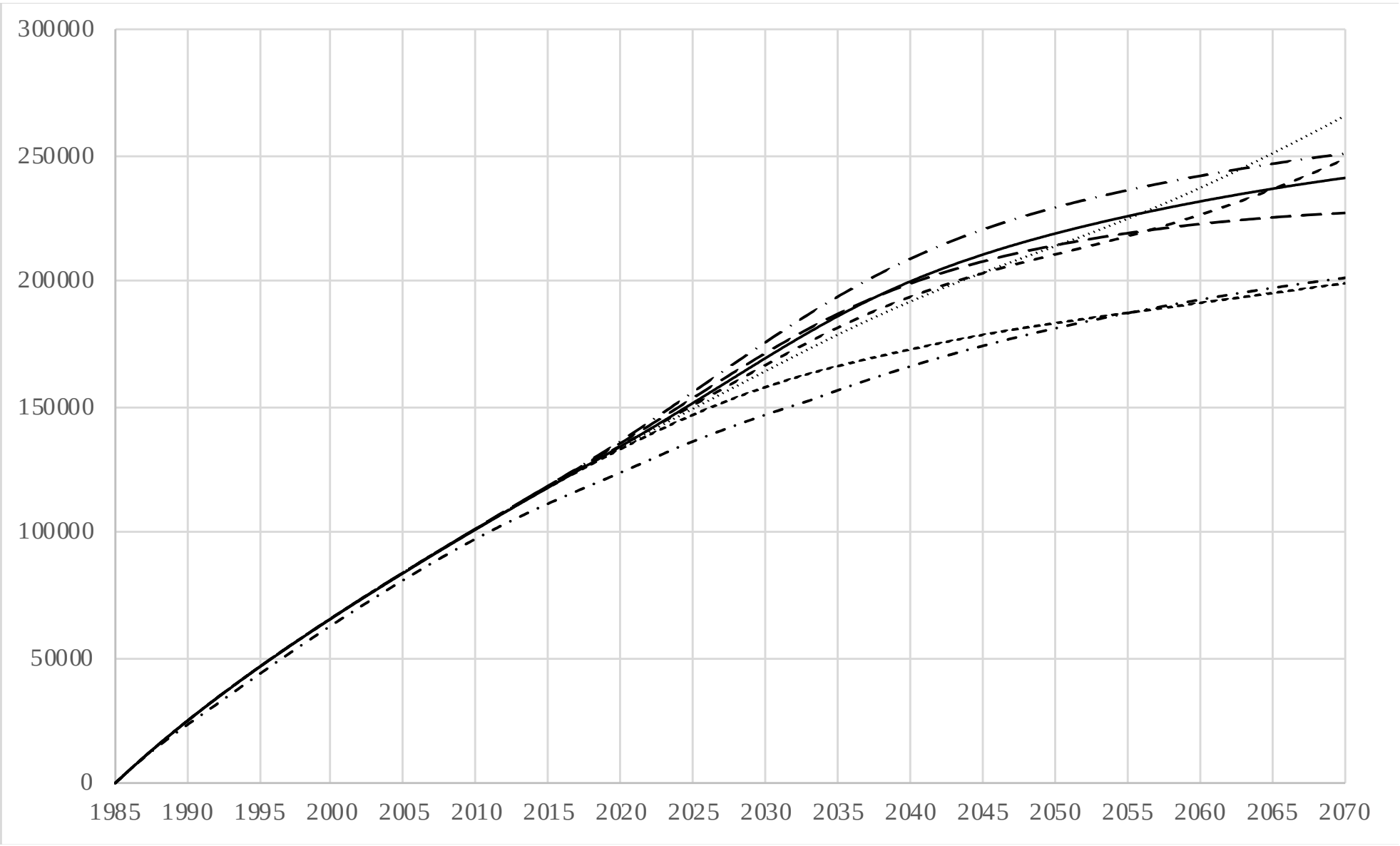

(b) Cumulative GHG emissions (Mtons)

Baseline | Hybrid-incumbent | Niche-incumbent | Landscape pressure
Sociotechnical transition | Niche favoured relationships | Predator-prey

Figure 9. GHG emissions under seven scenarios

## 5. Discussion

The overall results revealed that HEV as a hybrid technology established a symbiotic relationship on the technological development side with both ICEV and BEV in the business-as-usual situation. This implies that hybrid technologies may not necessarily act as predator towards incumbent technologies and at the same time as prey towards emerging technologies on the technological development dimensions. The symbiotic relationship on the technological development side depicts a bundle of positive externalities in the form of spillforwards and spillbacks among emerging, hybrid and incumbent technologies. HEV provided more positive externalities for the technology development side of BEV than for ICEV, and this means that the technological development spillforwards from a hybrid technology towards an emerging technology is likely to outweigh its technological development spillbacks towards an incumbent technology. This may be why BEV's technological development dimensions declined in the absence of HEV in niche-incumbent scenario. Additionally, results showed that hybrid technologies may show an explorative behaviour on the technological development side, and that this HEV's explorative behaviour was associated with the higher growth of BEV's technological development dimensions. This kind of explorative behaviour on technology development side aligns with the definition of hybrid technologies as exploration-hybrid innovations observed by Gatti Junior et al. (2021). This implies that by building exploration-hybrid innovations manufacturers can enhance their explorative capability on technology development dimensions, navigate uncertainties and risks related to technology development and improve their competitiveness while developing a more favourable business environment.

On the market development side, such as vehicle sales, Google popularity, and market share, HEV established a symbiotic relationship with ICEV but a mixed mode of symbiosis, parasitism, commensalism, and amensalism with BEV. While BEV provided market development spillbacks to both ICEV and HEV it received either negative or no spillforwards from these powertrain systems. Additionally, for most of the time, HEV benefited from the growth of BEV's market development dimensions and at the same time inhibited the growth of BEV's market development dimensions via parasitic or amensalism relationships. On the one side, this implies that hybrid technologies may not necessarily act as preys towards emerging technologies on the market development dimensions because they seem more likely to inhibit the growth of emerging technologies' market development dimensions through predatory behaviour. On the other side, hybrid technologies may not necessarily act as predators towards incumbent technologies' market development side as they seem more likely to support their market development dimensions through symbiotic relationships. In parallel with this, an exploitative behaviour from HEV was observed on the market development side, which align with the definition of hybrid technologies as exploitation-hybrid innovations observed by Gatti Junior et al. (2021). This implies that by building exploration-hybrid innovations manufacturers can generally take on a more defensive or path-dependence innovation strategy to closely monitor the lock-in, core rigidities and radical innovations in the transition process while sustaining the current business environment.

The conclusion is that while the assumed transitionary and supportive role of hybrid technologies is fulfilled for technological development dimensions of emerging technologies, their predatory or inhibitive role is only partially fulfilled for both technological and market development dimensions of incumbent technologies. Accordingly, hybrid technologies are redefined as hybridised solutions that establish multimodal and multidimensional relationships with both incumbent and emerging technologies, and that may suspend or accelerate the transition process, depending on relationship modes and points of interaction. This also should be noted that expecting the genuine predator-prey behaviour assumed in the transition research may not necessary be the best long-term option. Results revealed that while the genuine predator-prey behaviour may lead to a higher market share and sales for BEV and at the same time a lower market share and sales for ICEV on the short term, it may not dramatically alleviate the level of GHG emissions in the long term. Instead, reinforcing the various socio-technical dimensions of BEV and at the same time weakening the various socio-technical dimensions of ICEV and HEV (as in sociotechnical scenario) promised as the best option both in short and long terms.

Finally, the findings of this research reject the illusion of the 'sailing-ship effect[3]' as incumbents' improvement reaction to the threat of emerging technologies (Adner & Snow, 2010; Mirzadeh Phirouzabadi et al., 2020a; Sick et al., 2016). Both versions of the sailing ship effect, i.e., 'try harder' and 'spillovers' were frequently observed in the results. The spillovers effect was observed in the form of spillbacks and spillforwards between the three powertrain systems. Incumbent manufacturers and suppliers can suspend the fully transition process for a longer time by allowing the various components of an emerging technology to spill over into their incumbent technology in the form of a hybrid technology(Schiavone, 2014). The 'try-harder' effect, which is the easiest and least risky approach that incumbent manufacturers can employ to extend the lifetime of their incumbent technology by making incremental improvements over time (Gilfillan, 1935; Schiavone, 2014; Sick et al., 2016), was also observed in the results, e.g., an increase in the technological development dimensions of ICEV within a more favourable and supporting environment for BEV in landscape pressure scenario (e.g., higher oil price and higher WTW cost). ICEV's market development dimensions also declined much slower in landscape pressure scenario than in baseline.

## 6. Acknowledgement

This research did not receive any specific grant from funding agencies in the public, commercial, or not-for-profit sectors.

[3] The sailing-ship effect itself originally refers to the innovation efforts that the incumbent sailing ships made after noticing the introduction of the newly developed steam ships in the 19th century (Gilfillan, 1935; Sick et al., 2016).